\documentclass[twocolumn,showpacs,preprintnumbers,amsthm,amsmath,amssymb]{revtex4}
\usepackage{epsfig,amsmath}
\usepackage{graphicx}
\usepackage{dcolumn}

\newcommand{\beq}{\begin{equation}}
\newcommand{\eeq}{\end{equation}}
\newcommand{\beqa}{\begin{eqnarray}}
\newcommand{\eeqa}{\end{eqnarray}}
\newcommand{\lam}{\lambda}
 \newcommand{\rh}{\rho}
\newcommand{\ga}{\gamma} 
 
 \newcommand{\si}{\sigma}
 \newcommand{\om}{\omega}

\def\jmo#1{{ J.\ Mod.\ Opt.} {\bf#1}}
\def\jpb#1{{ J.\ Phys.\ B} {\bf#1}}
\def\jpa#1{{ J.\ Phys.\ A} {\bf#1}}
\def\pra#1{{ Phys.\ Rev. A\/} {\bf#1}}
\def\prb#1{{ Phys.\ Rev. B\/} {\bf#1}}

\def\prl#1{{ Phys.\ Rev.\ Lett.} {\bf#1}}

\begin{document}


\title{ Many-Body Separability of Warm Qubits }

\author{Ting Yu}
\email{ting@pas.rochester.edu}

\author{J.\ H.\ Eberly}
\email{eberly@pas.rochester.edu} 

\affiliation{ Rochester Theory Center, and  Department of Physics and 
Astronomy, University of Rochester, New York 14627, USA }


\date{July 21, 2007}

\begin{abstract}
We analyze the separability of the joint state of a collection of 
two-level systems at finite temperature $T$. The fact that only 
separable states are found in the neighborhood of their thermal 
equilibrium state guarantees that unimpeded thermal decoherence will 
destroy any initially arranged entanglement in a finite time.
\end{abstract}


\maketitle


Significant advances have occurred in the last decade in both the 
mathematical characterization of entanglement and in its 
applications, but it  remains an important open issue how 
entanglement responds to the influence of environmental noise 
\cite{Nielsen-Chuang}. The interaction of a quantum system 
with its surroundings is a generic phenomenon because real physical 
systems cannot be isolated completely from external noise.

So far, it has been remarkably challenging to find anything generically provable about the process of decoherence (i.e., disentanglement). For example, if there are more than two bodies involved it is still not known how to calculate 
whether a general many-body mixed state is separable or not.  Here we provide many-body results concerning separability of a physically important class of mixed states. These arise from consideration of interactions with thermal reservoirs and properties of thermal equilibrium. In many respects these are the least specialized considerations that can be imagined, as every real physical system is exposed to a thermal bath at some non-zero temperature.

\begin{figure}[!t]
\includegraphics[width= 6 cm]{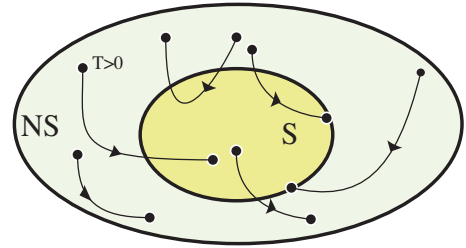}
\caption{{\footnotesize \label{fig.Trajectories3} Possible routes of 
dynamical evolution of states are suggested by the trajectories 
sketched in an imaginary non-metrical state space where the interior 
solid line is the boundary between separable (S) and non-separable 
(NS) states. The arrow heads simply indicate a direction of 
evolution. Our first result is that evolution to thermal steady state 
for $T > 0$ must end at a position on the interior of S, as shown for 
the route labelled $T > 0$.}}
\end{figure}

We suppose only that each sub-system in a many-body network of qubits 
relaxes over time to its thermal equilibrium state. We then 
demonstrate that the many-qubit thermal equilibrium state is special, 
in the sense that it is embedded in a finite neighborhood of 
completely separable states. So far as we know, the specification of 
a many-body state with a finite neighborhood of guaranteed 
separability has been established up to the present time only for the 
identity state (see Zyczkowski, et al. \cite{Zyczkowski-etal98}). Our 
result is easily extended to a wide range of other physically 
accessible states, essentially filling all of separable state space. 
This is indicated in Fig. \ref{fig.Trajectories3}

{\it Preliminary Considerations:--}  Our general result is established without speculating 
about approximate or ``reasonable" measures of many-body entanglement, and 
is first proven abstractly. We then provide new results 
regarding wide-scale entanglement decay in a network of two-level 
systems interacting with independent thermal reservoirs.

Our results are ultimately based on the fact that the thermal 
equilibrium state itself is separable and diagonal in the energy 
representation. We may represent it by the density matrix
\beq\label{rho_0}
\rh_0 = diag(p_1,p_2,...),
\eeq
with $p_j > 0,\ (j=1,2,..., N)$, for any finite temperature $T > 0$.

Although our main result is equally valid for an arbitrary number of subsystems,
it is convenient in the beginning to divide the many-body state space 
$\cal{H}$ into two parts and consider a bipartite system: $\cal{H} = 
{\cal H}^A \otimes {\cal H}^B$. We assume that both $\cal{H}^A$ and 
$\cal{H}^B$ are finite many-dimensional spaces. An operator $\rho$ 
acting on $\cal{H}$ describes a state if (1) $tr(\rho) = 1$, and (2) 
$tr(\rho P)\geq 0$ for any projection operator $P$. That is, $\rho$ 
is a positive Hermitian operator.

We will work from the fact  that a bipartite density matrix $\rho$ is 
non-separable (entangled) if and only if there exists a Hermitian 
``entanglement witness" matrix $W$ with the two properties \cite{backgroundtexts}:
\beqa \label{Wthm1}
&&tr(W\rho) < 0, \quad {\rm and} \\
\label{Wthm2}
&&tr(W\sigma) \geq 0,\quad {\rm for\ all\ separable\ states\ \sigma}.
\eeqa


{\it Finite Neighborhood of Separability:--}
We assert that for a density matrix with the thermal equilibrium form 
(\ref{rho_0}) there exists a neighborhood $U$ of $\rho_0$ such that 
all states in that neighborhood are separable. The result of 
Zyczkowski, et al. \cite{Zyczkowski-etal98} can be regarded as the 
limiting case of our result as $T \to \infty$. We will prove our 
assertion by showing that the opposite case must be false - i.e., the 
assumption that an entangled state exists in every neighborhood of 
$\rho_0$ leads to a contradiction.

First suppose that a neighborhood $U_n$ of $\rho_0$ has an entangled 
state $\rho_n$ in it. Then consider an infinite sequence of smaller 
neighborhoods, $U_{n+1} \subset  U_n$, all having at least one 
entangled state. This implies a sequence of entangled states $\rho_n$ 
converging to  $\rho_0$ as $n \to \infty$.

According to (\ref{Wthm1}) and (\ref{Wthm2}), for each entangled 
state $\rho_n$ there must exist a witness operator $W_n$ that 
satisfies both 
\beq
\label{1eq}
tr(W_n\rho_n) < 0,\quad {\rm and} \quad tr(W_n \sigma) \geq 0,
\eeq
for $\rho_n$ and all separable states $\sigma$.  Therefore, for the sequence of states and witnesses, we get (as $n \to \infty$):
$tr(W_{n}\rho_{n}) \to tr(\tilde{W}\rho_0),
tr(W_{n} \sigma) \to tr(\tilde{W}\sigma),$
for all separable states $\sigma$.  Moreover, by combining these 
expressions, and remembering the separable character of $\rho_0$, we 
have
\beqa \label{condition1}
tr(\tilde{W}\rho_0) \leq 0, \\
\label{condition2}
tr(\tilde{W}\sigma) \geq 0.
\eeqa
Then since $\rho_0$ is acceptable as a possible $\sigma$, we conclude 
that satisfying both (\ref{condition1}) and (\ref{condition2}) requires
\beq \label{zerosum}
tr(\tilde{W}\rho_0) = p_1\tilde{W}_{11} + p_2\tilde{W}_{22} +.... = 0.
\eeq

At this point a useful observation is that the state represented by 
$\sigma_1 = diag(1,0,..., 0)_A \otimes diag(1,0,...,0)_B$ is 
obviously separable, and if $\tilde W$ is an entanglement witness, 
then $tr(\tilde W\sigma_1) \geq 0$ requires $\tilde W_{11}\geq 0$, 
and by extension $\tilde W_{jj} \geq 0$ for all $j$.  Thus, since we 
are concentrating on the case in which the $p_{j}$s in (\ref{rho_0}) 
are all greater than zero, we conclude from (\ref{zerosum}) that all 
the diagonal elements $\tilde W_{jj}$ must vanish. 
Therefore, we have $tr\tilde{W} = tr(\tilde{W}I) = 0$, where $I$ is the 
identity matrix.  Since $\tilde{W}$ is not a zero matrix, there 
exists a product projection $P \otimes Q$ such that $tr(\tilde{W} P 
\otimes Q) \neq 0$  (see, e.g., \cite{proof}). By adding and 
subtracting $I$ and using $tr(\tilde{W}I) = 0$ again, this is 
sufficient to show that
\beq tr[\tilde{W}(P\otimes Q)] = -tr[\tilde{W}(I-P\otimes Q)] \neq 0.
\eeq
Both $P \otimes Q$ and $(I- P \otimes Q)$ are (at least proportional 
to) separable states, so (\ref{condition2}) cannot hold for both. 
This is the needed  contradiction that proves our main result.

An immediate implication of the existence of a separable neighborhood for a system consisting of $M$ qubits, each coupled to a local thermal heat bath, is that unimpeded thermal decoherence must destroy the entanglement of every initial state in a finite time, an effect we discussed for zero-temperature two-atom spontaneous emission in a previous note (\cite{Yu-Eberly04}, referred to below as YE for short). This has been labelled ESD for early-stage decoherence or ``entanglement sudden death", indicating that in order to reach $\rho_0$ asymptotically the system must enter the separable neighborhood of $\rho_0$ after only a finite time.


{\it $M$-Qubit Systems Under Thermal Noise:--}
The dynamics of pure thermal decoherence is  completely determined by the reduced density matrix of the M-qubit system, obtained  by tracing over the other qubits and the thermal reservoir's variables. In the familiar Born-Markov approximation, when each qubit is in contact with a broadband harmonic reservoir at temperature $T$, one finds a compact  Lindblad master equation (e.g., see \cite{Agarwal74}) for $M$ qubits ($\hbar=1$):
\beq \label{masterequ}
\frac{d}{dt}\rh = -i[H_{\rm sys}, \rho]+ \mathcal{L}(\rho),
\eeq
where $H_{\rm sys}=\sum_{i=1}^M\frac{1}{2}\om_i\si^{(i)}_z$ 
and  the Liouvillian superoperator is $\mathcal{L}(\rh) =
\sum_{i=1}^M \sum_{j=1}^4 (C^\dag_{ij}\rho C_{ij} -\frac{1}{2}\rho C_{ij} C_{ij}^\dag
-\frac{1}{2}C_{ij} C_{ij}^\dag \rho)$ and the Lindblad operators are $C_{i1} =
\sqrt{(\bar{n}+1)\Gamma} \sigma_-,\ C_{i2} =
\sqrt{\bar{n}\Gamma}\sigma_+ ,\ C_{i3} = \sqrt{(\bar{n}+1)\Gamma}\sigma_-,\  C_{i4} =\sqrt{\bar{n}\Gamma} \sigma_+.$ Here $\bar{n}$ is
the mean number of thermal reservoir quanta: $\bar{n} = 
1/(e^{\omega/k_BT} - 1)$.

The general solutions of the master equation (\ref{masterequ}) can
be expressed compactly by the Kraus operators $K_j$:
\beq \label{kraussolution}
\rho(t)=\sum_{i_1... i_M=1}^4  K_{i_1}\otimes...\otimes K_{i_M}  \rho(0) [K_{i_1}
\otimes ...\otimes K_{i_M}]^\dag,
\eeq
where the elementary Kraus operators $K_j$ with $\sum_{j=1}^4 
K_{j}^\dag K_j = I$ are the same for each qubit, given equal 
temperatures for all the reservoirs, and are explicitly given by
\begin{eqnarray}
       \label{k1}
K_1&=&\sqrt{\frac{\bar
n+1}{2\bar{n}+1}}\left(\begin{array}{clcr}
\gamma(t) & 0\\
0 &  1\\
\end{array}
        \right),\\
        K_2&=&\sqrt{\frac{\bar n +1}{2\bar{n}+1}}\left(
\begin{array}{clcr}
0    &   0 \\
\om(t) & 0\\
\end{array}
        \right),\\
K_3&=&\sqrt{\frac{\bar n}{2\bar{n}+1}}
\left(\begin{array}{clcr}
1  &   0 \\
0  &  \ga(t)\\
\end{array}
        \right),\\
K_4 &=&\sqrt{\frac{\bar n}{2\bar{n}+1}}
        \left(\begin{array}{clcr}
0  &  \om(t) \\
0  &   0 \\
\end{array}
        \right), \label{k4}
        \end{eqnarray}
where the time-dependent Kraus matrix elements are
$\gamma(t)= \exp\left[-\frac{1}{2}\Gamma(2\bar n + 1)t\right],
\om(t)=\sqrt{1-\gamma^2(t)}. $  The operators $K_1, K_2$ provide the transitions
from the excited state $|+\rangle$ to the ground state $|-\rangle$
caused by stimulated and spontaneous emission whereas the operators 
$K_3, K_4$ account for absorptive transitions from
the ground state $|-\rangle$ to the excited state $|+\rangle$.  

For the $M$-qubit system described by the master equation above, the steady state will be given by $(\ref{rho_0}$) . Let $\rho(t)$ denote the state of $M$ qubits at $t$. Then  $\rho(t) \rightarrow \rho_0$ as $t \rightarrow \infty$. Our generic result implies that a finite ESD time, $t_{\rm esd}$, exists for an arbitrary initial state. A simple expression for $t_{\rm esd}$ is given in YE for two-atom spontaneous emission.


{\it Pair Entanglement in Thermal States:--} For a two-qubit subsystem, our generic results established above can be realized in a more concrete way by explicitly solving the time evolution.  Bipartite dynamics under thermal noise for $T>0$ has been treated previously in considering various aspects of entanglement, e.g., generation \cite{Kim-etal02}, fragility \cite{Yu-Eberly02, Doll}, influence of squeezed reservoirs \cite{Daffer-etal03, Ikram-etal07, Guo07}, Brownian particle diffusion \cite{Dodd-Halliwell04}, and universality of ESD \cite{TerraCunha07, AlQasimi-James07}. We now show explicitly that when each  qubit relaxes to its equilibrium state asymptotically, the entanglement between any pair of qubits will always terminate in a finite time irrespective of the initial states of the qubits.

For any qubit pair, the steady state will be the diagonal matrix (\ref{rho_0}) with $N=4$:
\begin{equation}
\label{sol} \rho_{\rm st} = diag(p_1,p_2,p_3,p_4),
\end{equation}
in the standard basis $|++\rangle,|+-\rangle,|-+\rangle,|--\rangle$, 
where the thermal probabilities are $p_1 = \bar{n}^2/(2\bar{n}+1)^2,\
p_2 = p_3 = \bar{n}(\bar{n}+1)/(2\bar{n}+1)^2,\ p_4 =
(\bar{n}+1)^2/(2\bar{n}+1)^2.$

A standard measure of entanglement for our two-qubit system  is 
Wootters' concurrence \cite{Wootters}, denoted $C$.  By construction, 
the concurrence varies from $C=0$ for a separable state to $C=1$ for 
a maximally entangled state. For the two qubits $A$ and $B$ we have:
\beqa \label{Lambda}
&&C^{AB}(\rho) = \max\left[0,\Lambda(\rho) \right],\quad{\rm where}\\
&&\Lambda({\rho}) \equiv \sqrt{\lam_1} - \sqrt{\lam_2} - \sqrt{\lam_3} -
\sqrt{\lam_4},
\eeqa
and the $\lam_i$ are the eigenvalues in decreasing order of the 
matrix $\zeta_\rho \equiv \rho(\sigma^A_y\otimes \sigma^B_y) \rho^{*} 
(\sigma^A_y\otimes \sigma^B_y)$. Here $\rh^{*}$ denotes the complex 
conjugation of $\rh$ in the standard basis given above and 
$\si^{A,B}_y$ are the Pauli matrices expressed in the same basis.

Formulated in this way,  a necessary and sufficient condition for 
$\rho$ to be separable (zero entanglement) is $\Lambda \leq 
0$. We have noted previously \cite{Yu-Eberly07} that the negative 
values of $\Lambda$ possess quantum state information not 
contained in the concurrence $C(\rho)$. Both $\Lambda = 0$ and 
$\Lambda < 0$ signal that the two-qubit density matrix $\rho$ is 
separable, and when $\Lambda(\rho) < 0$, we call the state $\rho$ a 
super-separable state. The distinction for quantum state trajectories 
is shown in Fig. \ref{fig.Trajectories3}, where some trajectories 
terminate inside the S zone ($\Lambda < 0$) and some terminate 
exactly at the edge of the S zone ($\Lambda = 0$).  For the thermal 
steady state $\rho_{\rm st}$, straightforward calculations yield 
$\Lambda({\rho_{\rm st}}) = -2\sqrt{p_2p_3} = 
-2\bar{n}(\bar{n}+1)/(2\bar{n}+1)^2 \leq 0$, so the steady state is 
separable for any reservoir temperature including zero, but 
super-separable for all $T > 0$.

\begin{figure}[!b]
\includegraphics[height=3.3 cm]{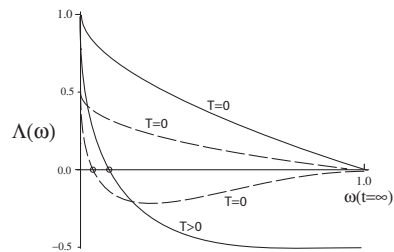}
\caption{{\footnotesize \label{fig.LambdaPlots} Four types of disentanglement trajectory are plotted via their $\Lambda$ values, where the horizontal axis covers $t = [0, \infty]$, scaled according to 
$\omega(t)$ given below Eqn. (\ref{k4}). The solid lines show 
evolution for $T=0$ and $T>0$ and each is appropriate to both
Bell states, which start with $\Lambda = 1$. While ESD is universal for $T > 0$, the dashed lines, starting from $\Lambda = 1/2$, show evolution of mixed states and demonstrate that ESD can also occur for $T = 0$. The dots where two curves cross the $\omega$ axis identify $t_{\rm esd}$ times.}}
\end{figure}

In order for the system to reach the zone of super-separable states, 
$\Lambda$ must go from positive to negative values. Since $\Lambda$ 
is a real number, this means it must cross the value $\Lambda = 0$, 
and this suggests a different graphical representation of 
trajectories with metrical elements that can assist interpretation, 
as in Fig. \ref{fig.LambdaPlots}. Whenever the long-time steady state 
of a two-qubit system is super-separable, the system must experience ESD. This means that it must become separable after only a finite time, as in the lower two curves in the figure, where the dots identify two locations for $t_{\rm esd}$. This possibility was apparently first noticed in several different physical contexts independently (see  \cite{Rajagopal-Rendell01, Zyczkowski-etal02, Diosi03, Daffer-etal03, Dodd-Halliwell04, Yu-Eberly04}). A two-qubit ESD time was evaluated in YE, and ESD was first reported experimentally by the Davidovich group \cite{Almeida-etal07} and also recently by the Kimble group in atomic ensembles \cite{Kimble07}.

In some cases of interest the ESD times $t_{\rm esd}$
can be determined explicitly. It is instructive to follow 
entanglement evolution for the two Bell states $|\Psi_{\pm}\rangle = 
\frac{1}{\sqrt{2}}(|+-\rangle \pm |-+\rangle)$.  From 
(\ref{kraussolution}), the evolving Bell density matrix at $t$ is 
given by
\begin{equation}
\label{sol1} \rho(t) = \left[
\begin{array}{clcr}
a(t) & 0  &  0 & 0 \\
0  & b(t) & z(t) & 0 \\
0  & z(t) & c(t) & 0\\
0  &  0 & 0 & d(t)
\end{array} \right],
\end{equation}
where the time-dependent matrix elements are given by the
following:
$a(t)=p_2\om^2+p_3\ga^2\om^2,
   b(t)=c(t)=p_1\ga^2 +p_2\ga^2+p_3(\ga^4 +1  + \om^4),
d(t)=p_1\om^2 +p_3\ga^2\om^2,
   z(t)=\pm \frac{1}{2}\gamma^2.$
Then it is straightforward to compute $\Lambda(\rh(t)) =
2|z(t)|-2\sqrt{a(t)d(t)}$. Thus, we conclude that when
\beq
t \geq t_{\rm esd} = \frac{1}{\Gamma} 
\ln\left[\frac{1+2\sqrt{p_1p_2}}{2\sqrt{p_1p_2}}\right],
\eeq
the initial Bell states $|\Psi_\pm\rangle$ become completely
disentangled.  As suggested by the two solid curves in 
Fig.~\ref{fig.LambdaPlots}, the higher the temperature the shorter 
the disentanglement decay time $t_{\rm esd}$. The upper solid curve 
is consistent with YE, showing no Bell disentanglement in a finite time 
when the environments are at zero temperature.


{\em Conclusion:--}
The central element of this note is our demonstration of a finite neighborhood of separable states surrounding the thermal equilibrium state, but the proof could equally well have been given for any state in a finite dimensional space having  diagonal form and positive definite elements. In fact continuously many other states of this type are also physically realizable, in some cases easily. Any completely incoherent partially excited state would suit, and an obvious example is the state associated in spin resonance with finite but ``negative" temperature or in laser physics with a positive partial inversion.

All of these also have finite neighborhoods of separability and 
all such neighborhoods  are clearly linked to each other. It is an open and interesting question whether the volume of this set of states can make up a significant or even finite fraction of the total separable volume. We note that our demonstration did not require knowing the finite number $M$ of qubits under consideration, and so the existence of a finite separable neighborhood is independent of $M$. The result is topological, and does not conflict either with the remark of Zyczkowski, {\it et al.}, \cite{Zyczkowski-etal98} that the volume of separable states decreases with system size, or the proof by Eisert, {\it et al.}, \cite{Eisert-etal02} that non-separable states are dense in continuously infinite Hilbert spaces.

Not surprisingly, ESD and other forms of dissipative entanglement dynamics and their control are of current interest, with examinations reported of many different dynamical processes and realizations of qubits, including coupled mechanical oscillators \cite{Rajagopal-Rendell01}, qubits in a spin chain and coupled to an Ising chain \cite{Pratt-Eberly01,Sun-etal07},
multi-cavity QED \cite{Yonac-etal07,Vaglica}, 
spin ensembles coupled via lossy photonic channels \cite{Madsen-Mollmer07},
qubit-qutrit combination \cite{Ann-Jaeger07}, two-qubit decoherence dynamics \cite{Privman-etal05,CuiHT-etal07,Hu06,Lastra,Adel} 
and multiple noises \cite{Yu-Eberly06}, 
to name a few.

To summarize, the main purpose of this note is to extend the study of entanglement in thermal and thermal-like environments to general multipartite systems with arbitrary initial states. We have achieved this goal by exploring topological properties of the state space of a many-body system. As an illustration, we have demonstrated the necessary thermal decoherence of two-qubit systems. The suppression of thermal decoherence and treatment of interacting qubits will be considered in future publications.

We acknowledge financial support from NSF Grant PHY-0601804 and
ARO Grant W911NF-05-1-0543.


\begin{thebibliography}{99}
\bibitem{proof} If the Hermitian operator $A$ satisfies $tr(A
P^A\otimes Q^B)= 0$ for any product projections  $P^A\otimes 
Q^B$, then $A=0$.
\end{thebibliography}
\end{document}